\documentclass[aps,superscriptaddress,showpacs,nofootinbib,eqsecnum]{revtex4}
\usepackage{amsmath,lscape,epsfig} \usepackage{amsfonts} 
\usepackage{amssymb,color,ulem} \usepackage{graphicx} \usepackage{multirow}
\usepackage{subfig}

\begin{document}

\title{ Anisotropy in the EoS of Magnetized Quark Matter}

\author{D\'{e}bora P. Menezes}
\affiliation{Depto de F\'{\i}sica - CFM - Universidade Federal de Santa
Catarina  Florian\'opolis - SC  - CEP 88.040 - 900 - Brazil}
\author{Marcus B. Pinto}
\affiliation{Depto de F\'{\i}sica - CFM - Universidade Federal de Santa
Catarina  Florian\'opolis - SC -  CEP 88.040 - 900 - Brazil}
\author{Constan\c ca Provid\^encia}
\affiliation{CFisUC, Department of Physics,
University of Coimbra, P 3004-516  Coimbra, Portugal}

\begin{abstract}
The anisotropies in the pressure obtained from the energy-momentum
tensor are studied for magnetized  quark matter within the  su(3)
Nambu-Jona-Lasinio model for both $\beta$-equilibrium matter and
quark matter with equal quark chemical potentials. The  effect of the magnetic field on the
particle polarization, magnetization and quark matter constituents is
discussed. It is shown that the  onset of the $s$-quark after chiral
symmetry restoration of the $u$ and $d$-quarks gives rise to a special
      effect on the magnetization in the corresponding density range:  a quite small
magnetization just before the $s$ onset is followed by a strong increase of
this quantity as soon as the $s$ quark sets in. It is also demonstrated that for
$B<10^{18}$ G within the two scenarios discussed, always considering a
constant magnetic field, the two components of pressure are practically coincident.

\end{abstract}


\maketitle

The structure of the QCD phase diagram is of utmost importance in
understanding many physical aspects of nature, ranging from the early universe
to possible nuclear liquid-gas and hadronic quark  matter phase
transitions to the physics of compact objects \cite {reviews}. Early
analyzes performed within the  Nambu--Jona-Lasinio model (NJL)
framework indicate that when strongly interacting  matter is subject
to intense magnetic fields the QCD phase diagram boundaries are
modified \cite{qcd+b}. Some of the most important changes concern the
size and location of the first order chiral transition region  since the
results show that a strong magnetic field favors this type of
transition. At the same time, at low temperatures, the value of the
coexistence chemical potential decreases as $B$ increases in
accordance with the inverse magnetic catalysis phenomenon (ICM) 
\cite {preis}.

 The low-$T$/high-$\mu$ region where a first order type
transition is expected to occur  is currently unavailable to lattice
QCD evaluations (LQCD). However, the region high-$T$/low-$\mu$,  has
already  been exploited using LQCD simulations which indicate, in
accordance with most model predictions, that the crossover observed
at $B=0$ persists when $B \ne 0$ \cite{earlylattice,preussker,lattice1,lattice2}. 
On the other hand, a major disagreement between recent LQCD results \cite{lattice1,lattice2} and model calculations regards the dependence of crossover pseudocritical temperature, $T_{\rm pc}$, on the strength $B$ of the magnetic field. Specifically, the lattice results 
of Refs.~\cite{lattice1,lattice2}, performed with $2+1$ quark flavors and physical pion 
mass values, predict an inverse catalysis,  with $T_{\rm pc}$ decreasing with $B$, 
while effective models predict an increase of $T_{\rm pc}$ with $B$. This problem has been recently addressed by different groups \cite{teoriaIMC, catalysis-ours} which basically agree that the different results stem from the fact that most effective models miss back reaction effects (the indirect interaction of gluons and $B$) as well as the QCD asymptotic freedom phenomenon.
At the same time, other important 
aspects of the effects of strong magnetic fields on the QCD phase
diagram have already been studied including the behavior of the coexistence
chemical potential and the location of the critical end point (CEP)
\cite{CEP1}, the dependence of the CEP on  strangeness, isospin and
charge asymmetry \cite{CEP2} and also the internal structure of the phase diagram
\cite{structure}.

Regarding physical observables, the understanding of magnetized quark matter  is particularly important
at low densities and high temperatures, which is the relevant regime for the present  heavy ion collisions experiments
\cite{kharzeev}, as well as at low temperatures and high densities,
which is the regime concerning magnetars \cite{magnetars}. 

As far as heavy ion-collisions are concerned the presence of a strong magnetic field
most certainly plays a role  despite the fact that,   in principle,
the   field intensity should decrease very rapidly  
lasting for about 1-2 fm/c only \cite {kharzeev}. The possibility that
this short time interval may    
\cite{tuchin} or may not \cite{mclerran} be affected by conductivity
remains under dispute. The effects of strong 
magnetic fields and their
relation with the impact parameters have also been discussed
\cite{event} while the particle yields dependence on a constant external
magnetic field has been investigated in a naive approach
\cite{paoli}. Another aspect related to the presence of strong
magnetic fields at the early stages of the collisions 
 is the anisotropy of photon production in heavy ion
collisions at the RHIC energies \cite{photon_anisotropy}.
The new PHENIX data brings some doubts on the conventional
picture of thermalization and subsequent hydrodynamics or infers the
possibility that a new photon production mechanism is possible.
These works tell us that there is much to be done if a
complete understanding  of the effects of hadronic matter subject to strong 
magnetic fields is expected. 

When we look at the recent literature on magnetars, the controversy is
already present at the level of calculating the energy-momentum
tensor. While some of the first works advocated that the pressure,
having in mind the thermodynamical pressure obtained from the
thermodynamical potential that relates pressure to density,
should be isotropic \cite{prc1,prc2, luizlaercio, rudiney}, based on an interpretation given 
in Ref. \cite{blandford}, 
other works were based on the fact that  the energy-momentum tensor
gives different contributions for the parallel and perpendicular
pressure \cite{ferrer2010,sedrakian, jorge, aurora} (see Ref. \cite {gregor} for a discussion based on LQCD). If two different pressures are
indeed present in the system, the usual way of using the equation of
state as input to the Tolman-Oppenheimer-Volkoff equations (TOV)  \cite{tov},
which determine the structure of a spherically symmetric body of isotropic material in static gravitational equilibrium,
to obtain compact objects macroscopic properties, as radii and masses,
has to be done with care. What is normally done is to observe at what
value of the magnetic fields, the two pressures start to deviate and
then use the TOV equations up to this strength, so that in principle,
the EoS used as input is {\it practically} isotropic \cite{jorge,
  aurora,veronica}. Another important aspect is related to the
contribution of the electromagnetic interaction to the pressure(s) and
energy density, a term proportional do $B^2$, where $B$ is the
magnetic field strength. Since no field larger then $10^{16}$G has been observed at the
  surface of a magnetar, but according to the virial theorem one could
  expect fields as strong as $\sim 10^{18}$G in the interior,
 an  {\it ad~hoc} exponential density-dependent
magnetic field was proposed in ref.~\cite{Pal} and widely adopted in 
subsequent works~\cite{Mao,Rabhi, prc2, rudiney,
Ryu, Rabhi2, Mallick, luizlaercio,Dex, njlv,Mallick2,Ro}.  This {\it
ansatz}, however,  violates Maxwell equations.  Another similar prescription for an {\it energy density
dependent} magnetic field was proposed in Ref. \cite{chaotic}. 

To avoid the use of the TOV equations, the authors of Ref. \cite{Mallick2}
treated the anisotropic pressure as a perturbation in a way similar to the Hartle-Thorn method,
generally used for slowly rotating neutron stars. In a more complete
treatment, the authors consider the anisotropy in solving Einstein's field equations
in axisymmetric regime with a fully general relativistic
formalism~\cite{Bocquet,Cardall,Micaela}. In both cases, once the
macroscopic properties are obtained, a small increase of the maximum
mass is found in contrast with the other previous works.

According to classical books on gravitation \cite{Misner, Zel}, when
anisotropies are present, the concept of pressure is not so well
defined. Based on the concepts discussed in these two books, in
Ref. \cite{chaotic} a small-scale chaotic field is used and the stress
tensor is modified, so that the resulting EoS is also isotropic. A curious
outcome is that the increase in the maximum stellar mass is also very
small, as found in Ref. \cite{Mallick2, Micaela}. Hence, it is clear that there is no unique way of computing magnetic
field effects on compact stars.

An estimation of the maximum magnetic field intensity supported
  by a star before magnetic field stresses give rise to the formation
  of a black hole may be obtained equating the magnetic field energy
  of an uniform field in a sphere with the star radius $R$
  to the gravitational binding energy. A maximum field of the order of
  $10^{18}$ G is obtained in agreement with  the maximum fields obtained in the
 framework of a relativistic magneto-hydrostatic formalism, of
the order of $\sim 5\times 10^{18}$G  ˙ with a nucleonic EOS  \cite{Cardall}, or  $\sim 3 \times
10^{18}$ G in Ref. \cite{broderick2002} with an hyperonic
EOS.  It was suggested that a  disordered field with $\langle B^2\rangle> \langle \vec
B\rangle^2$ could possibly give rise to larger fields in still stable
stars.  The previous estimations referred to stars that are bound by
gravitation. For self-bound stars larger fields could in principle
exist \cite{noronha2007,ferrer2010}. Taking these numbers as indicative we next
consider fields $B\le 1.5 \times 10^{19}$ G.

In the present work  we investigate, within the su(3) version of the Nambu-
Jona-Lasinio \cite{njl} model the quark matter polarization and magnetization, the thermodynamical
pressure, and  the parallel and perpendicular
pressure contributions obtained from the  energy-momentum tensor. We
consider both $\beta$-equilibrium matter and quark matter with
equal chemical potentials for the three flavors, { which we call
 symmetric quark matter or matter with isochemical potentials throughout the text.}
The first scenario
applies to neutron stars while the second is relevant to heavy ion collision investigations.
Some of these quantities are inputs for numerical codes that calculate the
structure of neutron stars subject to strong magnetic fields. 

The paper is organized as follows: In Sections I and II, the general
  formalism and the resulting equation of state, developed in
  Refs. \cite{prc1,prc2} are revisited. In Section III, the expressions for
  the magnetization and the anisotropic pressures are shown, with some
  of the details given in Appendix A. In Section IV the results are
  shown and discussed and in Section IV the final conclusions are drawn.

 \section{General formalism}

In order to consider { both symmetric quark matter and stellar quark matter} in $\beta$ equilibrium with
strong magnetic fields we introduce   the following Lagrangian density
\begin{equation}
{\cal L} = {\cal L}_{f}+{\cal L}_{l} - \frac {1}{4}F_{\mu
\nu}F^{\mu \nu} \,,
\end{equation}
{ which contains a quark sector, ${\cal L}_f$, a leptonic sector,
  ${\cal L}_l$, and the electromagnetic contribution.}
The quark sector is described by the  su(3) version of the
Nambu--Jona-Lasinio model
\begin{equation}
{\cal L}_f = {\bar{\psi}}_f \left[\gamma_\mu\left(i\partial^{\mu}
- {\hat q}_f A^{\mu} \right)-
{\hat m}_c \right ] \psi_f ~+~ {\cal L}_{sym}~+~{\cal L}_{det}~.
\label{njl}
\end{equation}
The ${\cal L}_{sym}$ and ${\cal L}_{det}$ terms are given by:
\begin{equation}
{\cal L}_{sym}~=~ G \sum_{a=0}^8 \left [({\bar \psi}_f \lambda_ a \psi_f)^2 + ({\bar \psi}_f i\gamma_5 \lambda_a
 \psi_f)^2 \right ]  ~,
\label{lsym}
\end{equation}
\begin{equation}
{\cal L}_{det}~=~-K \left \{ {\rm det}_f \left [ {\bar \psi}_f(1+\gamma_5) \psi_f \right] + 
 {\rm det}_f \left [ {\bar \psi}_f(1-\gamma_5) \psi_f \right] \right \} ~,
\label{ldet}
\end{equation}
where $\psi_f = (u,d,s)^T$ represents a quark field with three flavors, ${\hat m}_c= {\rm diag}_f (m_u,m_d,m_s)$ with $m_u=m_d \ne m_s$ is the corresponding (current) mass matrix while ${\hat q}_f={\rm diag}(q_u,q_d,q_s)$ is the matrix that
represents the quark electric charges. In the same equation, $\lambda_0=\sqrt{2/3}I$  where
$I$ is the unit matrix in the three flavor space, and
$0<\lambda_a\le 8$ denote the Gell-Mann matrices. The   t'Hooft
interaction term (${\cal L}_{det}$)  represents a determinant in flavor space which, for
three flavor, gives a six-point interaction \cite {buballa} 
and ${\cal L}_{sym}$, which is symmetric
under global  $U(N_f)_L\times U(N_f)_R$  transformations and corresponds to a
four-point interaction in flavor space. 
The model is non renormalizable, and as a regularization scheme for  the
divergent ultraviolet integrals  we use a sharp cut-off $\Lambda$ in
three-momentum space. The parameters of the model, $\Lambda$, the coupling 
constants $G$ and $K$
and the current quark masses, $m_u$ and $m_s$, are determined  by fitting
$f_\pi$, $m_\pi$ , $m_K$ and $m_{\xi'}$ to their empirical values.
 We adopt the parametrization of the model proposed in \cite{hatsuda}:
$\Lambda = 631.4 \, {\rm MeV}$ , $m_u= m_d=\,  5.5\, {\rm MeV}$,
$m_s=\,  135.7\, {\rm MeV}$, $G \Lambda^2=1.835$ and $K \Lambda^5=9.29$.

The leptonic sector is described by
\begin{equation}
\mathcal{L}_l=\bar \psi_l\left[\gamma_\mu\left(i\partial^{\mu} - q_l A^{\mu}
\right) -m_l\right]\psi_l \,\,,
\label{lage}
\end{equation}
where $l=e,\mu$. One recognizes this sector as being represented by
the usual QED type of Lagrangian density. As  usual, $A_\mu$ and $F_{\mu \nu }=\partial
_{\mu }A_{\nu }-\partial _{\nu }A_{\mu }$ are used to account
for the external magnetic field. We are interested in a
  static and constant magnetic field in the $z$ direction and hence, 
we choose the gauge $A_\mu=\delta_{\mu 2} x_1 B$.

\section {The EoS}

We now need to evaluate the thermodynamical potential for the three flavor
quark sector, $\Omega_f$, which as usual can be written as 
\begin{equation}
\Omega_f = -P_f = {\cal E}_f - T {\cal S}_f - \sum_f \mu_f \rho_f,
\end{equation} 
 where $P_f$ represents the pressure, ${\cal E}_f$ the energy density, $T$ the temperature,
${\cal S}_f$ the entropy density,  $\mu_f$ the chemical potential, and $\rho_f$ the quark number density. A similar expression can be written for the leptonic sector.

The total pressure for three flavor in $\beta$ equilibrium is then given by
\begin{equation}
P(\mu_f,\mu_l,B)= P_f |_{M_f}+ P_l |_{m_l} \pm \frac{B^2}{2}
\,\,,
\end{equation}
where our notation means that $P_f$ is evaluated in terms of  the quark effective mass, $M_f$, which is
determined in a (nonperturbative) self consistent way while $P_l$ is
evaluated at the leptonic bare mass, $m_l$. 
The term  $B^2/2$, arises due to the  kinetic  term of the electromagnetic field, $F_{\mu \nu}F^{\mu \nu}/4$, in the
original Lagrangian density. { Within the formalism used in the present
  work, the sign of this term comes from the stress tensor and is
  shown in the section where we discuss anisotropy.} 
Remark also that in the sequel our results will be presented in terms
of (vacuum) subtracted pressures ($\Delta P$) such that 
$P_f=0$ at $\mu_f=0$ ($f=u,s,d$) and $P_l=0$ at $\mu_l=0$
($l=e,\mu$). With this normalization choice 
only the magnetic pressure, ($ \pm B^2/2)$,  survives at vanishing chemical potentials. 
Again, a similar expression can be written for the leptonic
sector, apart from the color index. The lepton masses are
$m_e=0.511 \, {\rm MeV}$ and  $m_\mu=105.66\, {\rm MeV}$.

In the mean field approximation the pressure can be written as
\begin{equation}
P_f = \theta_u+\theta_d+\theta_s 
-2G(\phi_u^2+\phi_d^2+\phi_s^2) + 4K \phi_u \phi_d \phi_s \,\,,
\end{equation}
where the free gas type of term is 
\begin{equation}
 \theta_f=-\frac{i}{2}  {\rm tr}  \int  \frac {d^4 p}{(2\pi)^4} \ln \left(-p^2 + M_f^2 \right ),
\end{equation}
while the scalar condensates, $\phi_f$ are given by
\begin{equation}
\phi_f= \langle {\bar \psi}_f \psi_f \rangle= -i  \int \frac {d^4 p}{(2\pi)^4} {\rm tr}\frac{1}{(\not \!
p - M_f+i\epsilon)}
\label{cond}
\end{equation}
According to standard Feynman rules for this model, all the traces 
are to be taken over color ($N_c=3$) and Dirac space, but not flavor. 

The effective quark masses can be obtained self consistently  from 
\begin{equation}
 M_i=m_i - 4 G \phi_i + 2K \phi_j \phi_k, 
 \label{mas}
\end{equation}
with $(i,j,k)$ being any permutation of $(u,d,s)$. So, to determine
the EOS for the su(3) NJL at finite density and in the presence of a
magnetic field  we need to know the condensates, $\phi_f$, as well as
the contribution from the gas of quasiparticles, $\theta_f$. 
Both quantities, which are related by $\phi_f \sim d \theta_f  /
dM_f$, have been evaluated with great detail in
\cite{prc1,prc2} so that here we just quote the results:

 \begin{equation}
\theta_f= \left (\theta^{vac}_f+\theta^{mag}_f + \theta^{med}_f \right )_{M_f}\,\,,
\label{pressBmu2}
\end{equation}
where the vacuum contribution reads
\begin{equation}
\theta^{vac}_{f}=- \frac{N_c }{8\pi^2} \left \{ M_f^4 \ln \left [
    \frac{(\Lambda+ \epsilon_\Lambda)}{M_f} \right ]
 - \epsilon_\Lambda \, \Lambda\left(\Lambda^2 +  \epsilon_\Lambda^2 \right ) \right \},
\end{equation}
and where we have also defined $\epsilon_\Lambda=\sqrt{\Lambda^2 + M_f^2}$ with
$\Lambda$ representing a non covariant ultraviolet cut off,
the finite  magnetic contribution is given by
\begin{equation}
\theta^{mag}_f= \frac {N_c (|q_f| B)^2}{2 \pi^2} \left [ \zeta^{\prime}(-1,x_f) -  \frac {1}{2}( x_f^2 - x_f) \ln x_f +\frac {x_f^2}{4} \right ]\,\,,
\end{equation}
with   $x_f = M_f^2/(2 |q_f| B)$ while $\zeta^{\prime}(-1,x_f)= d
\zeta(z,x_f)/dz|_{z=-1}$ where $\zeta(z,x_f)$ is the Riemann-Hurwitz
zeta function. To take further derivatives, as well as for numerical purposes, it is useful to use the following  representation for this quantity  
\begin{equation}
\zeta^{\prime}(-1,x_f)=\zeta^{\prime}(-1,0)+\frac{x_f}{2}[x_f-1-\ln(2\pi) + \psi^{(-2)}(x_f)] \;,
\end{equation}
where $\psi^{(m)}(x_f)$ is  the $m$-th polygamma function and the $x_f$ independent constant is $\zeta^{\prime}(-1,0)=-1/12$.
The medium contribution can be written as
\begin{equation}
\theta^{med}_f=T \sum_{k=0}^{k_{f,max}} \alpha_k \frac {|q_f| B N_c }{4 \pi^2}
 \int_{-\infty}^{+\infty} dp \left[ \ln\left(1 + \exp[-(E^*_f - \mu_f)/T]\right)
 +  \ln\left(1 + \exp[-(E^*_f + \mu_f)/T]\right) \right],
\label{PmuB}
\end{equation}
with $\alpha_0=1,\,\alpha_{k>0}=2$. In the above equation we have defined the energy dispersion $$E^*_f= \sqrt{p^2 + s_f(k,B)^2} ~ , \qquad  s_f(k,B)= \sqrt {M_f^2 + 2 |q_f| B k}\;.$$
When considering just the zero temperature case, Eq.(\ref{PmuB}) becomes:
\begin{equation}
\theta^{med}_f=\sum_{k=0}^{k_{f,max}} \alpha_k\frac {|q_f| B N_c }{4 \pi^2}  \left [ \mu_f \sqrt{\mu_f^2 - s_f(k,B)^2} -
 s_f(k,B)^2 \ln \left ( \frac { \mu_f +\sqrt{\mu_f^2 -
s_f(k,B)^2}} {s_f(k,B)} \right ) \right ] ,
\label{PmuBt0}
\end{equation}
where  $s_f(k,B) = \sqrt {M_f^2 + 2 |q_f| B k}$.
At $T=0$, the  upper Landau level (or the nearest integer)  is defined by
\begin{equation}
k_{f, max} = \frac {\mu_f^2 -M_f^2}{2 |q_f|B}= \frac{p_{f,F}^2}{2|q_f|B}.
\label{landaulevels}
\end{equation}

The condensates $\phi_f$ entering the quark pressure  at finite
density and in the presence of an external magnetic field can
be written as 

\begin{equation}
\phi_f=(\phi_f^{vac}+\phi_f^{mag}+\phi_f^{med})_{M_f}
\end{equation}
where

\begin{eqnarray}
\phi_f^{vac} &=& -\frac{ M_f N_c }{2\pi^2} \left [
\Lambda \epsilon_\Lambda -
 {M_f^2}
\ln \left ( \frac{\Lambda+ \epsilon_\Lambda}{{M_f }} \right ) \right ]\,\,,
\end{eqnarray}

\begin{equation}
\phi_f^{mag}
= -\frac{ M_f |q_f| B N_c }{2\pi^2}\left [ \ln \Gamma(x_f)  -\frac {1}{2} \ln (2\pi) +x_f -\frac{1}{2} \left ( 2 x_f-1 \right )\ln (x_f) \right ] \,\,,
\end{equation}
and
\begin{eqnarray}
\phi_f^{med}&=&
\sum_{k=0}^{k_{f,max}} \alpha_k \frac{ M_f |q_f| B N_c }{4 \pi^2}
\int_{-\infty}^{+\infty} dp \frac{(f_+ + f_-)}{E^*_f}\,,\label{MmuB}
\end{eqnarray}
where the Fermi distribution functions are
\begin{equation}
f_{\pm}= {1}/\{1+\exp[ (E^*_f \mp \mu_f)/T]\}\;.
\label{dfq}
\end{equation}
The quark density reads
\begin{equation}
\rho_f= \sum_{k=0}^{k_{f,max}} \alpha_k \frac{|q_f| B N_c }{2 \pi^2} \int_{-\infty}^{+\infty} dp (f_+ - f_-)
\label{densq}.
\end{equation}

At $T=0$, Eqs.(\ref{MmuB}) and (\ref{densq}) become:
\begin{equation}
\phi_f^{med}=
\sum_{k=0}^{k_{f,max}} \alpha_k \frac{ M_f |q_f| B N_c }{2 \pi^2}\left [\ln \left ( \frac { \mu_ f +\sqrt{\mu_f^2 -
s_f(k,B)^2}} {s_f(k,B)} \right ) \right ]\,\,,
\label{MmuBt0}
\end{equation}
and
\begin{equation}
\rho_f=
\sum_{k=0}^{k_{f,max}} \alpha_k \frac{|q_f| B N_c }{2 \pi^2} k_{F,f} \,\,,
\end{equation}
where  $k_{F,f}=\sqrt{\mu_f^2 - s_f(k,B)^2}$.
The entropy density ${\cal S}_f=-(\partial \Omega/\partial T)$ is
\begin{equation}
{\cal S}_f =-\sum_f \sum_{k=0}^{k_{f,max}} \alpha_k \frac {|q_f| B N_c }{4 \pi^2} \int_{-\infty}^{+\infty} dp
\left[f_+ \ln\left(f_+\right)
+\left(1-f_+\right) \ln\left(1-f_+\right)+(f_+\leftrightarrow f_-)
\right]\;.
\end{equation}
The corresponding leptonic contributions can be trivially obtained from the above quantities by replacing $M_f \to m_l$, $|q_f| \to |q_l|$ and $\mu_f \to \mu_l$. Since the leptonic masses are unaffected by strong interactions one considers their bare values which do not depend on $T$, $\mu$ and $B$ as opposed to the effective $M_f$ masses.
Therefore, the only piece which effectively contributes to the subtracted  pressure, defined as $\Delta P= P(T,\mu,B) - P(0,0,B)$, is  
\begin{eqnarray}
P_l^{med}= T\sum_{k=0}^{k_{f,max}} \alpha_k \frac {|q_l| B}{4 \pi^2}
 \int_{-\infty}^{+\infty} dp \left[ \ln\left(1 + \exp[-(E_l - \mu_l)/T]\right)
 +  \ln\left(1 + \exp[-(E_l + \mu_l)/T]\right) \right]\;,
\label{Pl}
\end{eqnarray}
from which the leptonic density can be written as
\begin{equation}
\rho_l= \sum_{k=0}^{k_{f,max}} \alpha_k \frac{|q_l| B}{2 \pi^2} \int_{-\infty}^{+\infty} dp (l_+ - l_-)
\,\,,
\end{equation}
where $\mu_l$ represents the leptonic chemical potential. The quantities $E_l$ and  $l_\pm$ can be obtained from their quark counterparts by the  replacements already mentioned. 

At vanishing temperatures the  above expressions {become}:

\begin{equation}
P_l^{med}=\sum_{l=e}^\mu \sum_{k=0}^{k_{l,max}} \alpha_k\frac {|q_l| B }{4 \pi^2}   \left [ \mu_l \sqrt{\mu_l^2 - s_l(k,B)^2} -s_l(k,B)^2 \ln \left ( \frac { \mu_l +\sqrt{\mu_l^2 - s_l(k,B)^2}}
    {s_l(k,B)}   \right ) \right ]~.\nonumber \\
\end{equation}
and
\begin{equation}
\rho_l =  \sum_{k=0}^{k_{l, max}}\alpha_k \frac{ |q_l| B }{2 \pi^2}   k_{F,l}(k,s_l) \,\,,
\label{rholmuB}
\end{equation}
where $k_{F,l}(k,s_l) =\sqrt{\mu_l^2 - s_l(k,B)^2}$.

\section{ The anisotropy in the pressure }

The parallel and the perpendicular components of the pressure can be
written in terms of the magnetization, ${\cal M} = {\partial \Delta
  P}/{\partial B}$, as \cite{blandford, veronica,Mallick}:

\begin{equation}
P_{\parallel}=\Delta P - \frac{B^2}{2} \quad {\rm and} \quad P_{\perp}=\Delta P -{\cal
  M}B + \frac{B^2}{2} \,
\label{eq:press}
\end{equation}
where $\Delta P$ stands for the already defined subtracted pressure.
{For a magnetic field in the $z$ direction, the
stress tensor has the form: diag$(B^2/2,B^2/2,-B^2/2)$ and this
explains the difference in sign appearing in the parallel and
perpendicular pressures.}
For the leptonic sector one easily gets 
\begin{equation}
\frac{ dP_l^{ med}}{dB}=\frac{P_l^{med}}{B}- \frac{ |q_l|B}{4\pi^2} \sum_{k=0}^{k_{f,max}} \alpha_k(k|q_l|) \int_{-\infty}^{+\infty} dp \frac{1}{E_l}[l_+ + l_-] \;,
\end{equation}
which, at $T=0$ becomes:
\begin{equation}
\frac{d P^{ med}_l}{dB}= \frac{P_l^{med}}{B}- \frac{B |q_l|}{2\pi^2} \sum_{k=0}^{k_{max}} \alpha_k \ln \left ( \frac{ \mu_l + \sqrt{\mu_l^2-s_l^2}}{s_f} \right ) (k|q_l|) \;,
\end{equation}
 whereas for the quark sector one obtains

\begin{equation}
\frac{d P_f}{d B} = \theta_u^\prime+\theta_d^\prime+\theta_s^\prime 
-4G(\phi_u \phi_u^\prime+\phi_d \phi_d^\prime+\phi_s \phi_s^\prime) + 4K (\phi_u^\prime \phi_d \phi_s +\phi_u \phi_d^\prime \phi_s+ \phi_u \phi_d \phi_s^\prime) \,\,,
\end{equation}
where
\begin{equation}
 \theta^\prime_f=(\theta^{\prime\, vac}_f+\theta^{\prime \, mag}_f+\theta^{\prime \, med}_f)_{M_f} \,\,\,,
\end{equation}
and
\begin{equation}
 \phi^\prime_f=(\phi^{\prime \, vac}_f+\phi^{\prime \, mag}_f+ \phi^{\prime \, med}_f)_{M_f} \,\,\,.
\end{equation}
The primes denote derivatives with respect to $B$ and the explicit
form of each term can be found in the appendix A. 

\section{Results and discussion}

In this section we present and discuss results concerning
several properties of $\beta$-equilibrium quark matter and symmetric
quark matter (with equal chemical potentials).  We first discuss some
general properties of quark matter, in particular, its particle content
and particle polarization.  We then investigate how the
magnetization of quark matter changes with the magnetic field
intensity and finally we discuss the parallel and perpendicular pressure
contributions obtained from the energy-momentum tensor.

The quark spin polarization $\Delta$ is usually defined as
\begin{equation}
\Delta_i=\frac{\rho_i(\uparrow)-\rho_i(\downarrow)}
{\rho_i(\uparrow)+\rho_i(\downarrow)}.
\quad i=u,d,s
\end{equation}

\begin{figure}[tbp]
\centering
\begin{tabular}{cc}
\subfloat[]{\includegraphics[width=0.5\linewidth]{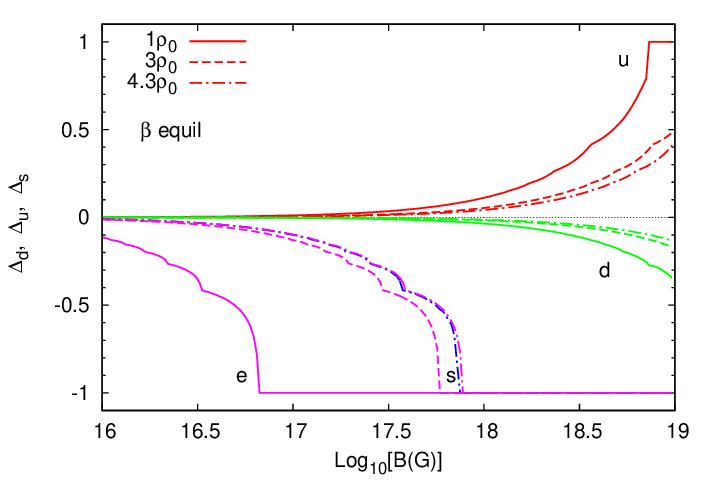}\label{fig1:a}}&
\subfloat[]{\includegraphics[width=0.5\linewidth]{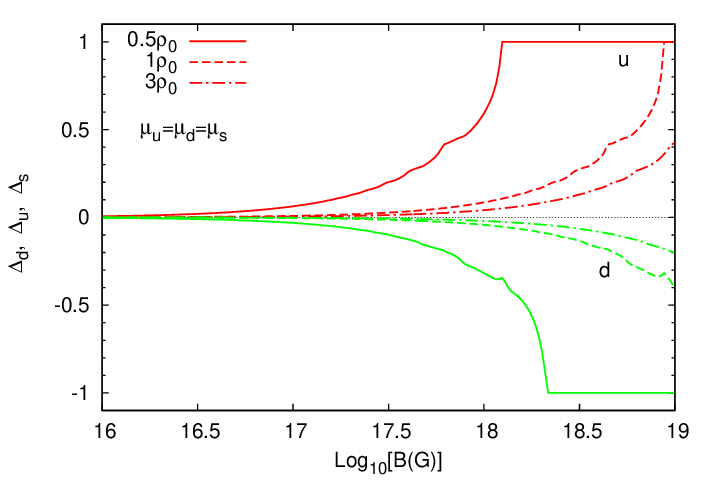}\label{fig1:b}}\\
\subfloat[]{\includegraphics[width=0.5\linewidth]{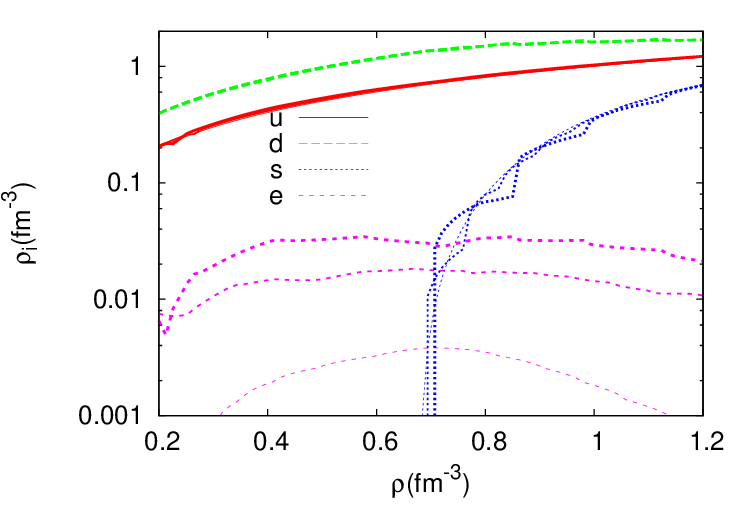}\label{fig1:c}}&
\subfloat[]{\includegraphics[width=0.5\linewidth]{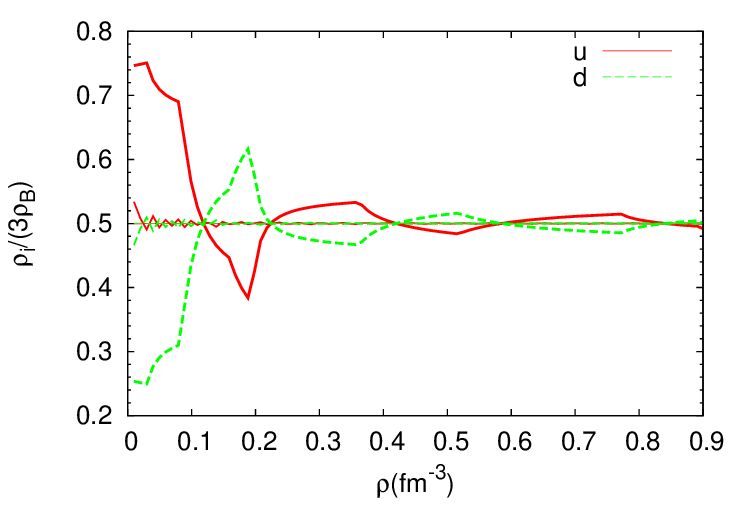}\label{fig1:d}}\\
\subfloat[]{\includegraphics[width=0.5\linewidth]{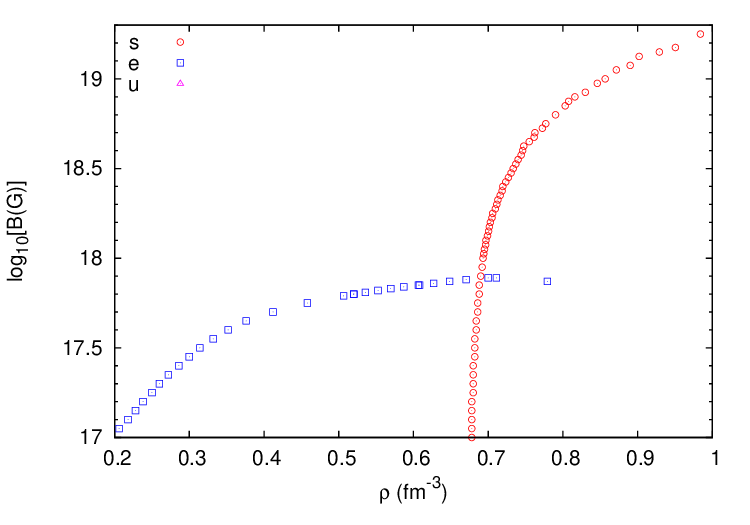}\label{fig1:e}}&
\subfloat[]{\includegraphics[width=0.5\linewidth]{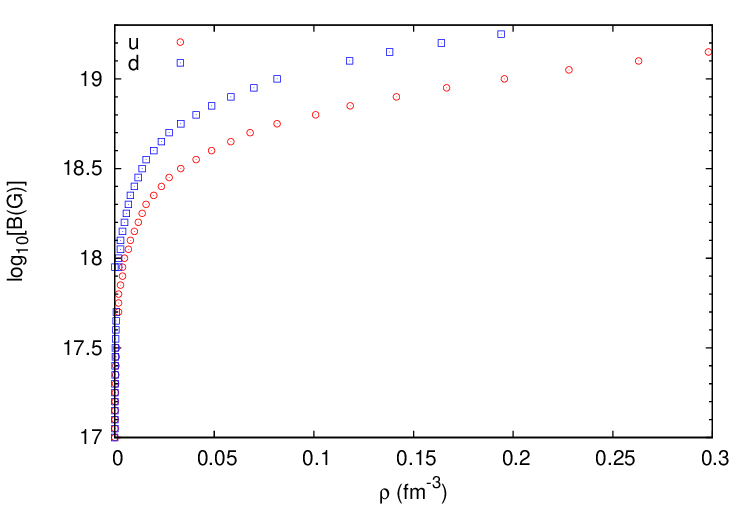}\label{fig1:f}}\\
\end{tabular}
\caption{$\beta$-equilibrium quark matter (left panels) and symmetric
  quark matter (right panels): a) and b) particle polarizations for several densities defined
  in terms of the saturation density $\rho_0=0.16$ fm$^{-3}$,  c) and d)
  particle {densities, absolute in c) and relative in d),} for $10^{17}$G, $10^{18}$G and $10^{19}$G
  magnetic field intensities, respectively, from the thinner to the
  thicker lines; e) 
  total polarization onset  density for the  $s$ quark and the
  electron in $\beta$-equilibrium matter and f) for the $u$ and $d$
  quarks in symmetric matter. For the densities
  representing $u$ and $d$ quarks the total polarization occurs for stronger
  fields and in symmetric matter the
  $s$ quark sets in at larger densities.}
\label{fig1} 
\end{figure}

In Fig. \ref{fig1} it is shown how the magnetic field affects
$\beta$-equilibrium matter (left panels) and symmetric quark matter
(right panels), in particular, the quark and electron polarizations,
their densities and onset of total polarization.  The
$\beta$-equilibrium matter onset of the $s$ quark occurs just below 0.7 fm$^{-3}$,
see Fig. \ref{fig1:c},
and, therefore,  close to these densities the $s$ quark density is small
and feels strongly the magnetic field, total polarization being
attained with $B<10^{18}$ G. In denser matter, the larger densities of
$s$ quarks require larger magnetic field intensities for a total polarization.
In symmetric quark matter the $s$ quarks set in above 0.9
fm$^{-3}$, a density that presently is not attained in the laboratory.
 In $\beta$-equilibrium matter  $u$
and $d$ quarks  are not totally polarized for fields below 10$^{19}$
G, however in symmetric quark matter $u$ and $d$ quark total polarization
occurs at small densities that do not exist in quark
stellar matter with a surface baryonic density which is of the order of
 0.3 fm$^{-3}$.  In Fig. \ref{fig1:a} we also show information on the electron
polarization. The magnetic field increases the electron content and
for $10^{18}$G their density is practically constant and equal to $\sim
0.01$ fm$^{-3}$, see
 Fig. \ref{fig1:c}. At the onset of the $s$ quark the density of
 electron has always a maximum, above which the electron fraction
decreases. This means that electrons are totally polarized for fields
$B\gtrsim 9\times 10^{17}$G. 

For totally polarized matter, all particles lie on the lowest Landau level (LLL). In
this case, the dependence of the pressure on the magnetic field intensity  of a
gas of free particles occurs only
through a multiplicative factor that defines the LLL degeneracy, and
the magnetization  is independent of $B$. For the NJL model the
interaction terms give a nonlinear dependence to the pressure
even above for totally polarized matter. This contribution becomes
small when the chiral symmetry is restored. A comparison with the
polarization obtained in quark matter described by the MIT bag model
is shown in \cite{aurora1}, where it is also seen that the
 polarization of the system increases with the increase of the
magnetic field and the total polarization occurs for $B \simeq 2
\times 10^{19}$ G for matter in $\beta$-equilibrium.

\begin{figure}[tbp]
\centering
\begin{tabular}{cc}
\subfloat[]{\includegraphics[width=0.5\linewidth]{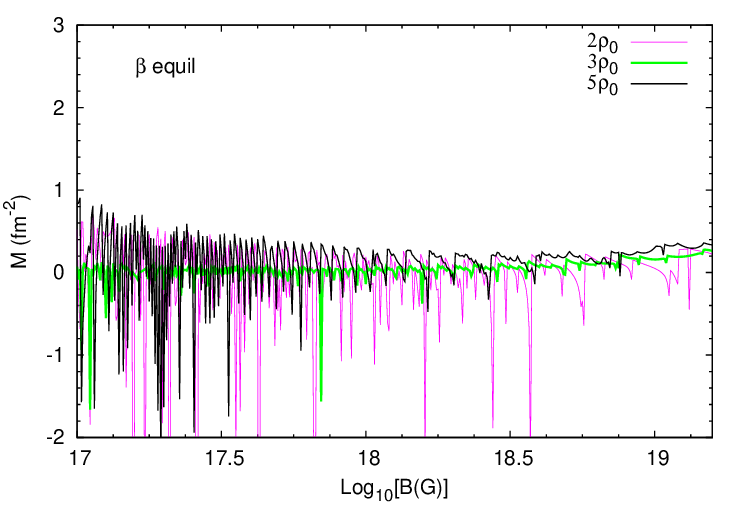} \label{fig2:a}}&
\subfloat[]{\includegraphics[width=0.5\linewidth]{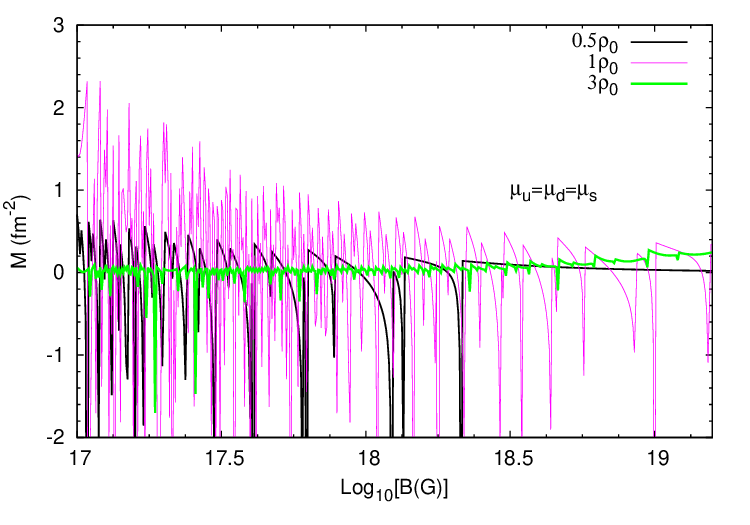} \label{fig2:b}}\\
\subfloat[]{\includegraphics[width=0.5\linewidth]{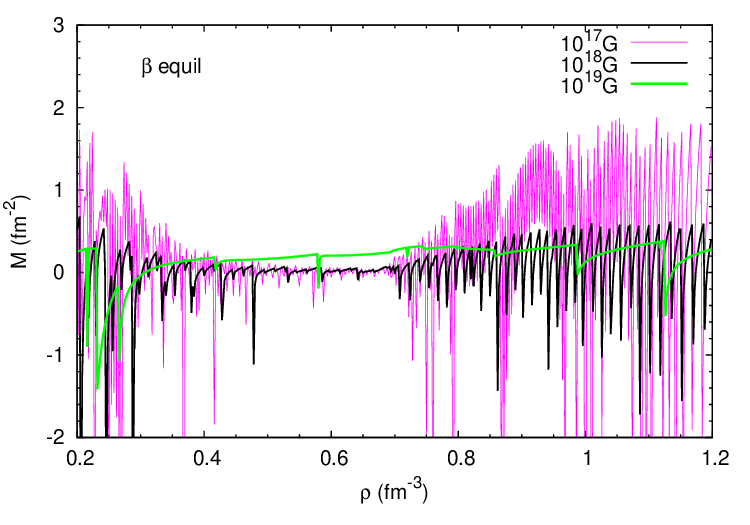} \label{fig2:c}}&
\subfloat[]{\includegraphics[width=0.5\linewidth]{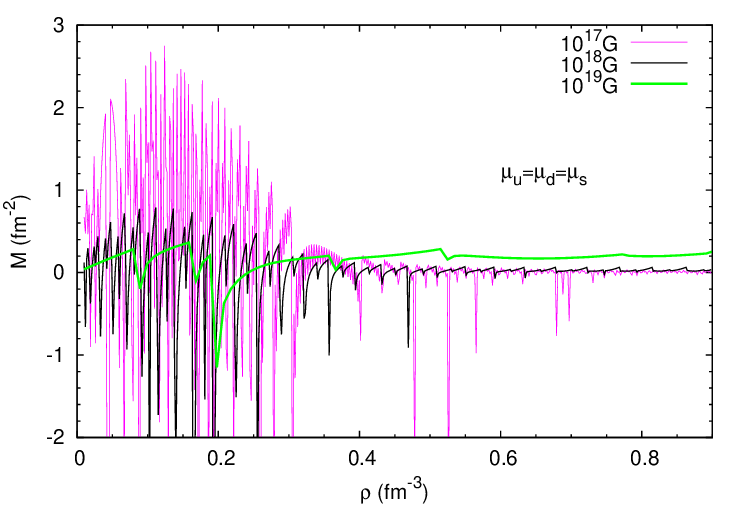} \label{fig2:d}}
\end{tabular}
\caption{ Magnetization for
 a) and c) $\beta$-equilibrium matter and b) and d) isochemical potential quark
 matter. Densities of interest for each scenario are considered. In a)
 and b) the magnetization is shown as a function of $B$ and in c) and
 d) as a function of the baryonic density.
\label{fig2} }
\end{figure}
In order to determine the two pressure contributions, the
magnetization is obtained using Eq. (\ref{eq:mag}). In Fig. \ref{fig2}
this quantity is plotted for $\beta$-equilibrium matter,
Figs. \ref{fig2:a} and  \ref{fig2:c}, and quark matter with equal chemical potentials,
Figs.  \ref{fig2:b} and \ref{fig2:d}. For each case we have considered a set of densities
of interest: a) the surface  baryonic density of a quark
star is  $\sim 2\rho_0$ and in the interior we may have densities
larger than
$5\rho_0$; b) in heavy ion collisions  we may have
densities below $\rho_0$ and  do not expect  densities as large as 5$\rho_0$.
The magnetization has a term that explodes whenever  $\mu_f=s_f$,
i.e., whenever $B$ approaches a $n\ne 0$ Landau level from below, see
Eq. (\ref{phimed}) and the corresponding discussion in Ref. \cite{sedrakian}. The
spikes in this figure occur precisely at this values of $B$. A small
number of spikes occurs if only a few LL are occupied, e.g. if the
magnetic field is very strong with respect to the Fermi momentum
of the particle. In Figs.  \ref{fig2:a}  and  \ref{fig2:b} for the
densities represented this occurs for the larger fields. 
 For the symmetric matter at smaller densities, fewer spikes are
obtained for larger magnetic fields.
 It is interesting to notice that for $\beta$-equilibrium matter there is a range of
densities between 0.5 and 0.7 fm$^{-3}$ where for all the fields shown
the number of oscillations is small: this occurs before the onset of
the $s$ quark and after the $u$ and $q$ quark restoration of chiral
symmetry. In Fig. \ref{fig2:c}, for densities lower than 0.3
  fm$^{-3}$ and in between 0.8 and 1.1 fm$^{-3}$, one can clearly sees
  that oscillations with high frequencies are modulated by smaller
  frequency oscillations. The same is seen in Fig. \ref{fig2:d}, for
  densities larger than 0.3 fm$^{-3}$.
This superposition of fluctuations with different frequencies is due to a mixing of
particles with different charges and masses. In $\beta$-equilibrium
matter the oscillations defined by the $u$ quark are wider apart
because the $u$ quark density is smaller and its charge larger, both
effects adding to a reduction of the number of occupied LL, see
Fig. \ref{fig2:c}. This same effect if also present in
Fig. \ref{fig2:d} where, for similar  $u$ and $d$ quark densities,
the charge of the $u$ explains the difference.
Our results differ substantially from the ones obtained in \cite{sedrakian} and
  \cite{aurora1} (apart from the choice of units), since the MIT bag
  model, used in the calculation of the quark magnetization in these
  references, do not present negative magnetization. Physically, this
  means that while the NJL model studied here presents paramagnetic
  (${\cal M} >0$) as well as  diamagnetic (${\cal M} < 0$) behavior
  while the MIT model, studied in Ref. \cite {veronica}, displays only
  a paramagnetic phase. This difference can be better understood by
  recalling that the NJL magnetization receives contributions from
  $\theta^{'med}_f$   and $\phi^{'med}_f$ while only the former
  contributes to the MIT model \cite {veronica}. As  remarked earlier
  $\theta_f$ represents the contribution from a gas of quasi-particles
and  $\phi_f$ represents the quark condensate. As
  Eq. (\ref{phimed}) reveals,  $\phi^{'med}_f$ may present divergencies leading to the oscillation between diamagnetic and paramagnetic states observed in our results   (before the LLL is reached). 

\begin{figure}[tbp]
\centering
\begin{tabular}{cc}
\subfloat[]{\includegraphics[width=0.5\linewidth]{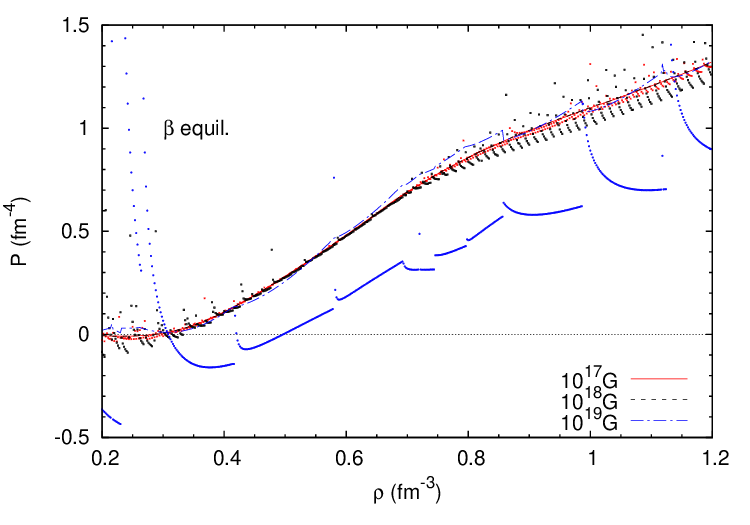} \label{fig3:a}}&
\subfloat[]{\includegraphics[width=0.5\linewidth]{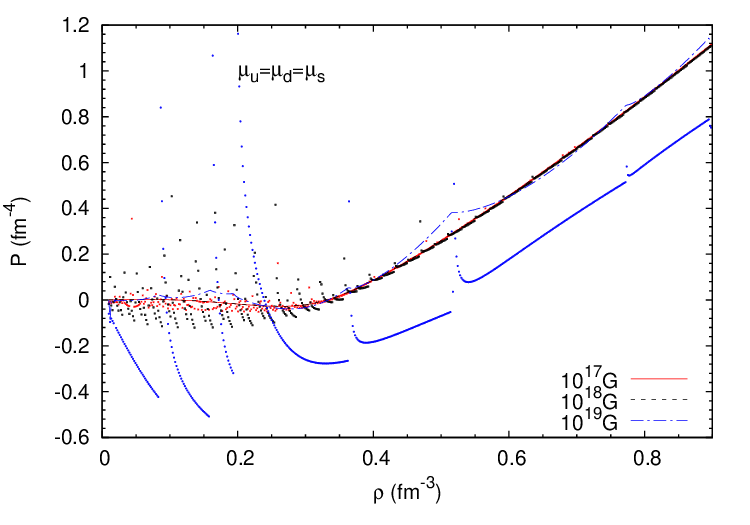} \label{fig3:b}}\\
\subfloat[]{\includegraphics[width=0.5\linewidth]{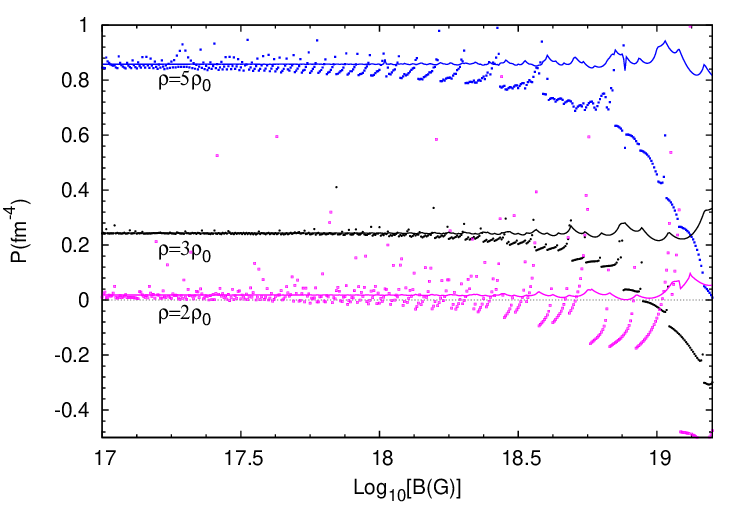} \label{fig3:c}}&
\subfloat[]{\includegraphics[width=0.5\linewidth]{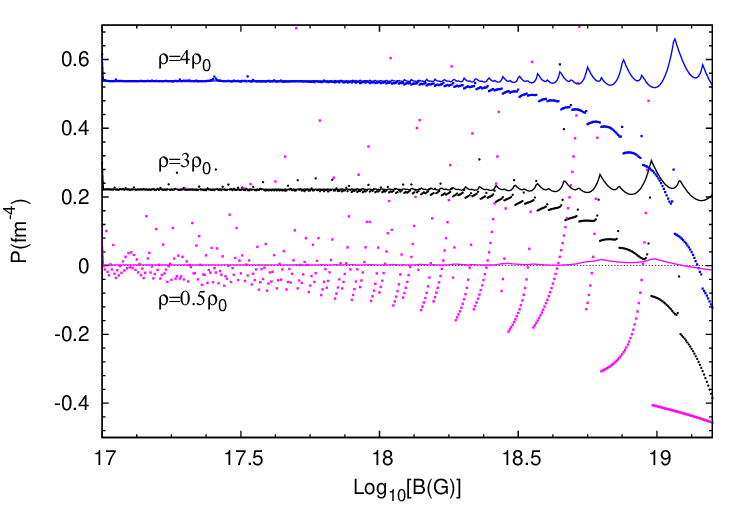} \label{fig3:d}}
\end{tabular}
\caption{Parallel (lines)  and perpendicular (marks) pressures for
 a) and c) $\beta$-equilibrium matter and b) and d) isochemical potential quark
 matter. Densities of interest for each scenario are considered. In a)
 and b) the pressures are shown as a function of the baryonic density
 and in c) and d) as a function of $B$ . For fields $B\le 10^{18}$ G
 the perpendicular press falls on top of the parallel contribution
 except for the van Alphen oscillations due to the magnetization contribution.
\label{fig3}}
\end{figure}

In Fig. \ref{fig3} the two pressure contributions defined in Eq. 
(\ref{eq:press}), except for the terms proportional to $B^2$, are
plotted for both $\beta$-equilibrium matter (left
panels) and symmetric matter (right panels). As in the previous
discussion for each scenario we take the same representative values of the
magnetic field intensity or baryonic density.
One should notice that when we analyze matter in
  $\beta$-equilibrium, we have in mind that quark stars have no crust,
  since the pressure is zero for a still finite density, 
as clearly seen  in Fig. 3 a). One of the effects of the magnetic
field could be a change of the density at the surface, 
although non-negligible effects occur only for magnetic fields above
$10 ^{18}$ G.  A mantle could have been added as was done
in \cite{glendenning95}. Moreover, in \cite{melrose} 
 it was shown that within the NJL a much larger electron chemical
 potential is expected at the surface allowing more easily for a
 mantle to exist. Understanding how this mantle could be affected in
 the presence of a magnetic field is an interesting question that,
 however, is beyond the scope of the present work. 

For fields $B< 10^{18}$ G both contributions, the parallel and the
perpendicular pressures, are practically coincident. However, due to
the magnetization contribution entering the perpendicular pressure,
discontinuities occur whenever a new LL level is reached, giving rise
to a series of discontinuities that would correspond to unstable
regions, since $d{\cal M}/dB<0$. 
{ Let us point out that in the region $0.5<\rho<0.7$ fm$^{-3}$, and
for $B<10^{18}$G, these discontinuous contributions to the
perpendicular pressure contribution are  practically   zero, because, as
seen before, the magnetization is very small in this range of densities. }
As compared with calculations performed
with the MIT bag model \cite{veronica}, the behavior is the same,
except that the deviation of the two pressures takes place at even
larger magnetic fields with the MIT bag model.  Nevertheless,
had we  considered not only the matter contribution but also the contributions from the 
the terms proportional to $B^2$, the deviation would start at lower
magnetic field intensities.
The exact value of the magnetic field where both pressures start
  to deviate depends on the model used and on the chosen fixed
  density (or corresponding chemical potential), as can be seen when we compare our
  results with the ones shown in \cite{sedrakian} and \cite{aurora1}, but
  the qualitative results in average are the same.

One should bear in mind that the discontinuities seen in Fig. \ref{fig3}
 are washed out at finite temperatures as the ones that we expect in heavy ion
collisions \cite{ours_finite}. As suggested in Refs. \cite{blandford,noronha2007,sedrakian} the
unstable regions may give rise to domains with disordered fields that
would allow to continuous phase transitions between regions with a
different number of LL. 

\begin{figure}[tbp]
\centering
\begin{tabular}{cc}
\subfloat[]{\includegraphics[width=0.5\linewidth]{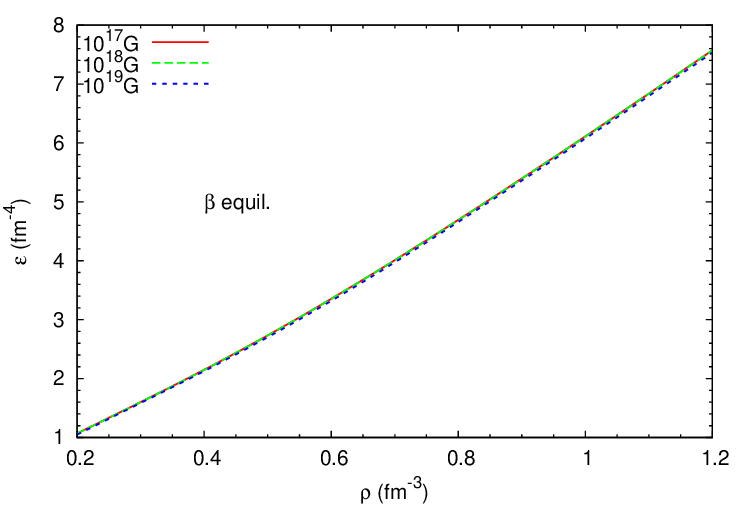} \label{fig4:a}}&
\subfloat[]{\includegraphics[width=0.5\linewidth]{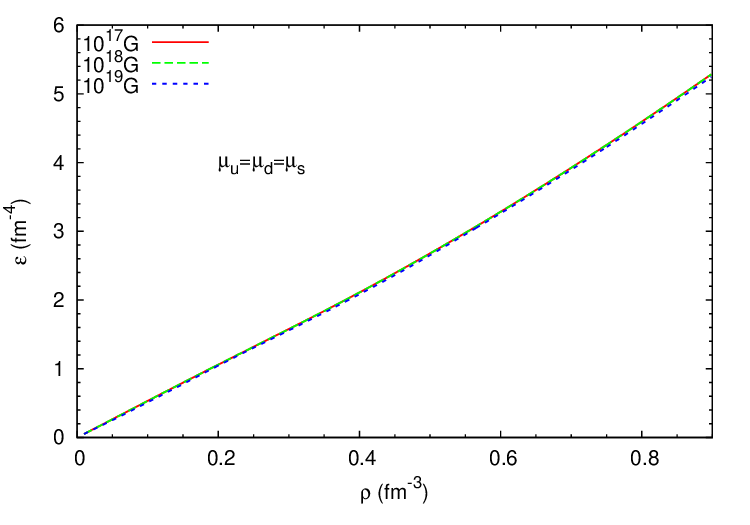} \label{fig4:b}}\\
\subfloat[]{\includegraphics[width=0.5\linewidth]{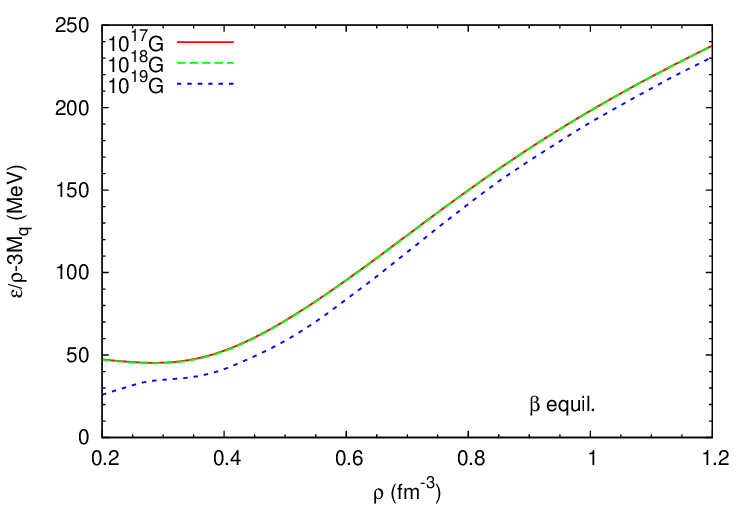} \label{fig4:c}}&
\subfloat[]{\includegraphics[width=0.5\linewidth]{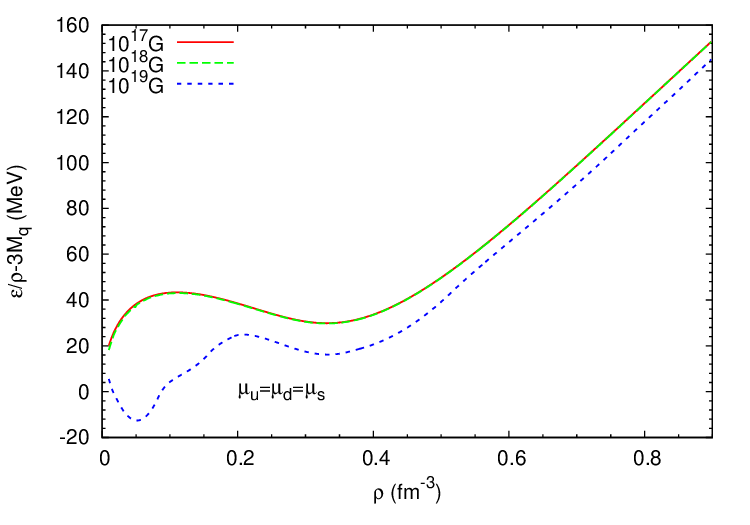} \label{fig4:d}}
\end{tabular}
\caption{a) and b) Energy density and c) and d) quark {\it binding energy}
  for matter in $\beta$-equilibrium in the left panels and symmetric
  matter in the right panels. 
\label{fig4}}
\end{figure}

 Finally, we plot in Fig. \ref{fig4}, top figures, the energy
  density as a function of the number density and in the figures at
  the bottom, a {\it quark binding energy} for the cases of matter in
  $\beta$-equilibrium (left panels) and isochemical potential matter
  (right panels). The energy density does not depend on the strength
  of the magnetic field, what corroborates the fact that thermodynamic
  consistency is achieved \cite{sedrakian} On the other hand, as
  already expected from previously published results \cite{prc1,prc2},
  the density per particle decreases with the increase of the magnetic
  field due to the change of the quark density with the magnetic
  field.

\section{Final remarks}

In this work we have examined the effects of strong magnetic fields in
quark matter, as described by the NJL model, paying special attention
to effects due to pressure anisotropy. Two scenarios were
investigated: matter in $\beta$-equilibrium, as the one possibly present in
quark stars or in the core of hybrid stars and symmetric matter with
isochemical potentials, as the one present in heavy ion
collisions. For the first case, large densities and magnetic fields up
to the order of $10^{18}$ G are of interest whereas for symmetric matter, the relevant
densities are somewhat smaller while the magnetic fields can be larger.

Part of our work has been devoted to the  analysis of quark matter constituents, the onset of $s$ quarks
and how much polarized matter can be present when different field intensities are considered since these are important ingredients for stellar modeling. 

Magnetization, which is  an important quantity for the description of magnetized matter, has also been considered in the present work and by investigating this quantity we conclude  that the amount of spikes, related to
the filling of the LL, depends quite substantially on the density and
on the scenario examined. Of particular interest is the fact that very
few spikes are seen for densities between 0.5 and 0.7 fm$^{-3}$ in
$\beta$-equilibrium matter, no matter the intensity of the magnetic
field, because the particles that constitute matter occupy only
  a few LL. For symmetric matter, the same pattern is found for densities
larger than 0.5 fm$^{-3}$, when the number of occupied LL are small
for the three different quarks.

This effect in $\beta$-equilibrium matter is due to the late onset of the $s$ quark, precisely
  for $\rho\gtrsim 0.7$fm$^{-3}$. Due to its large constituent mass
  magnetic fields satisfying $B\lesssim 10^{18}$G, as the ones
  possibly existing inside magnetars, give rise to the filling of many
  LL and large contributions to the magnetization and perpendicular
  pressure, in contrast to the behavior at densities just below the
  $s$ quark onset. This effect is clearly seen comparing the pressure
  for $\rho=3\rho_0$ corresponding to $u$ and $d$ quark matter with
  restored chiral symmetry with the pressure  obtained for either
  $\rho=2\rho_0$ or $\rho=5\rho_0$ as a function of $B$: in the first
  case chiral symmetry is still not completely restored and in the
  second the onset of $s$ quark has occurred.  If a quark model with
  complete chiral symmetry restored such as the MIT bag model, or a
  model without the strangeness degree of freedom such as the su(2)
  NJL is used to describe $\beta$-equilibrium quark matter, a
  different behavior will
  probably  occur
for $B\le 10^{18}$G and $\rho>0.5$ fm$^{-3}$ right until the star
center, e. g. $\rho\sim 1.2$ fm$^{-3}$ \cite{prc2,melrose} ,  in
particular,  both pressure contributions will be coincident.

 Therefore within the su(3) NJL model,  a non negligible effect of the magnetic field on the quark star structure close to the surface and in the interior is expected even for $B\sim 10^{18}$ G. Taking as reference the calculation done in \cite{rezzolla2012}  using  pure toroidal magnetic field equilibrium models of
relativistic stars for both non-rotating and rotating hadronic  stars,  fields as large as $10^{18}$ G were obtained  on the equatorial plane deep inside the star. In this calculation, however,  the magnetic field effects on the EOS have been neglected. 

We have then looked at the pressure anisotropy obtained from  the components
 of the energy-momentum tensor that
  define the parallel and the perpendicular pressure contributions and
examined the relation between parallel and perpendicular pressures for
both scenarios described above when the magnetic field is fixed and
also when the baryonic density is kept constant. We have observed that the
larger the magnetic field intensity, the larger the discontinuities in
the perpendicular pressure. It is important to note that the
magnetization is not responsible for the complete picture, because it
comes multiplied by the magnetic field in the calculation of the
perpendicular pressure. In agreement with Refs. \cite{jorge, veronica}
we have observed that when the  densities are fixed,
the parallel and perpendicular pressures are practically coincident up to very
large magnetic fields, what could justify the use of isotropic matter
hydrostatic equations
in stellar calculations. However, the integration of the full relativistic hydrostatic
equations still requires the inclusion of the magnetic field
contributions which involve a $B^2$ term that gives rise to a quite large
effect.

At this point, we  would like to emphasize that compact star macroscopic
  properties, as masses and radii were not computed because to trust
  our results  we would need either a self-consistent model, as the one developed in
  \cite{Bocquet,Cardall,Micaela} or a simpler approach as the one in
  \cite{Mallick2} that uses a more appropriate space geometry. 
In \cite{Micaela} it was shown that the contribution to the EOS  of
  magnetic fields of the order of $7\times 10^{17}$G are  negligible
  on the determination of the maximum mass  compared to the
  uncertainties existing among the EOS models. However, if we consider
  that a quark star is self bound, we may expect that  magnetic fields
  $\sim 10^{19}$ G or larger exist  in stable stars
  \cite{noronha2007,ferrer2010}, and, in this case, the effects of magnetic field
  on the EOS (see Fig. \ref{fig3}) are probably non negligible on the determination of the
  star structure.

One should also notice  that strong magnetic fields affect the proton and electron fractions of
$\beta$-equilibrium stellar matter. In particular it was shown that if
the electrons and protons are confined to their LLL, the Urca reactions
are open for an arbitrary proton concentration leading to a fast
neutron star cooling \cite{Leison99}, because the transversal momentum of
charged particles is defined only within an accuracy of $\Delta
p_\perp \sim \sqrt{eB}$. If this quantity is of the order of the
neutron Fermi momentum direct Urca processes are allowed. In
\cite{Baiko99}
is was discussed that even for much weaker fields there is already a
noticeable effect corresponding to a speeding up of the cooling near
the center of the star.
Recently, it was also shown that neutrino emissivity will be enhanced
due to B-induced unpairing of proton condensates 
\cite{sinha2015}.
 With quark matter the situation is
different. Unpaired quark matter may speedup the star cooling if the
pairing nucleon gap is large \cite{page2000}, but color
superconducting quark matter hinders neutrino emission.
However, if unpaired quark matter is considered under the effect of
strong magnetic fields it has been discussed that neutrino emission
may also be hindered because direct Urca will only occur under very
specific conditions \cite{Kouvaris2009}.
This effect requires that all particles are in their
  LLL. In the present calculation we have seen that for $B\lesssim10^{19}$ G
this is not the case, and therefore, within NJL the quark direct Urca
process will not be hindered for realistic magnetic field intensities.

Finally, we point out that other physical aspects, such as the fermion
anomalous magnetic moment, may also be important  
for a more realistic description when fields in excess of $10^{19}$ G are considered.
Indeed, the introduction of a magnetic field  induces a  new term
corresponding to the coupling between the field and the fermion
anomalous magnetic moment (AMM)  which may influence the EoS for very
high values of $B$.  In \cite{broderick00} 
it has been shown that the inclusion of the anomalous magnetic moment
contribution stiffens the stellar matter EOS if $B>10^{18}$G, and may
originate a total spin polarization of neutrons. For magnetic field
intensities below the critical intensity, obtained equating the
particle cyclotron energy to its rest mass, the Schwinger perturbative
determination of the anomalous magnetic moment, corresponding to the
lowest order correction to the magnetic moment obtained at tree level,
is valid. Recently, it has been shown that for quark matter the scale
for the perturbative approach is set by the constituent quark mass
\cite{chang2011}, and therefore, the effect of including an AMM should
also be considered when describing quark matter under the effect of
strong magnetic fields. With a mass about one third the nucleon mass,
the critical field will be approximately one order of magnitude
smaller the nucleon critical field, but still larger than the maximum
field expected inside a neutron or quark star. Taking into account the
AMM will certainly lift the degeneracy of Landau levels.

\acknowledgements
D.P.M. and M.B.P. are partially supported by Conselho Nacional de
Desenvolvimento Cient\'{\i}fico e Tecnol\'{o}gico (CNPq-Brazil) and  
by Funda\c c\~{a}o de Amparo \`{a} Pesquisa e Inova\c c\~{a}o do Estado de Santa Catarina (FAPESC-Brazil), under project 2716/2012. C.P. acknowledges financial support 
by Project No. PEst-OE/FIS/UI0405/2014
developed under the initiative QREN financed by the UE/FEDER through the
program COMPETE $-$ ``Programa Operacional Factores de
Competitividade.

\section{Appendix A - Derivatives with respect to $B$}

As already shown in the text, the magnetization is given by:

$${\cal M}=-\left(\frac{d\Omega}{dB}
  \right)_\mu=-\left(\frac{\partial\Omega}{\partial B}
  \right)_\mu- \sum_f\left(\frac{\partial\Omega}{\partial
      M_f}\right)_\mu
\frac{d M_f}{dB},$$
but in equilibrium
$$\left(\frac{\partial\Omega}{\partial M_f} \right)_\mu=0\;,$$
then 
\begin{equation}
M=-\left(\frac{\partial\Omega}{\partial B}
  \right)_\mu\;.
\label{eq:mag}
 \end{equation}
 Thus, in the following expressions we do not need to include  terms
 containing $\frac{d M_f}{dB}$ which also means 
that the vacuum contributions $\theta^{\prime \,vac}_f$ and $\phi^{\prime \,vac}_f$ trivially vanish.  Then,

\begin{equation}
\theta^{\prime \, mag}_f= 2 \frac{\theta^{mag}_f}{B} -
 \frac{N_c |q_f| B}{2\pi ^2} \frac{M_f^2}{2B}  \left[ \ln
 \Gamma(x_f)-\frac{1}{2} ln (2 \pi) + x_f - (x_f - \frac{1}{2}) \ln(x_f) \right]\; ,
 \end{equation}
and
\begin{equation}
\theta^{\prime \, med}_f=\frac{\theta_f^{med}}{B}- \frac{N_c
  |q_f|B}{4\pi^2} \sum_{k=0}^{k_{f,max}} \alpha_k k|q_f| \int_{-\infty}^{+\infty} dp \frac{1}{E_f^*}[f_+ + f_-]\,\,.
\end{equation}
For $T=0$ the above relation becomes:
\begin{equation}
\theta^{\prime \, med}_f= \frac{\theta_f^{med}}{B}- \frac{N_c B |q_f|}{2\pi^2} \sum_{k=0}^{k_{max}} \alpha_k \ln \left ( \frac{ \mu_f + \sqrt{\mu_f^2-s_f^2}}{s_f} \right ) k|q_f|\,\,.
\end{equation}
A straightforward evaluation yields
\begin{eqnarray}
\phi^{\prime \, mag}_f=\frac{\phi_f^{mag}}{B} + \frac{N_c}{4\pi^2} \frac{M_f^2}{2B} \left \{ |q_f| B +M_f^2 \left[ \psi^{(0)}(x_f) - \ln(x_f)\right] \right \}\,\,,
\end{eqnarray}
where $\psi^0(x_f)=\frac{\Gamma^{\prime}(x_f)}{\Gamma(x_f)}$ is the
digamma function. The in medium contribution reads

\begin{eqnarray}
\phi^{\prime \, med}_f&=& \frac{\phi_f^{med}}{B} - \frac{N_c
                          |q_f|B}{4\pi^2} \sum_{k=0}^{k_{f,max}
                          }\alpha_k k|q_f| \int_{-\infty}^{+\infty} dp \left \{\frac{f_+}{(E^*_f)^2}\left [
\frac{1}{E^*_f} +\frac{f_+}{T}\exp[(E_f^*-\mu_f)/T]\right ] \right .  \nonumber \\
&+& \left . (\mu \leftrightarrow -\mu \;\;'\;\; f_+ \leftrightarrow f_- ) \right \} \;,
\end{eqnarray}
which,  at $T=0$, can be written as
\begin{equation}
\phi^{\prime \, med}_f= \frac{\phi_f^{med}}{B} 
-\frac{  N_c |q_f|B}{2 \pi ^2}
\sum_{k=0}^{k_{f,max}} \alpha_k   \frac{
  \mu_f M_f k |q_f|}{s_f(k,B)^2 \sqrt{\mu_f^2-s_f(k,B)^2}}  .
\label{phimed}
\end{equation}

\end{document}


\title{ Erratum: Anisotropy in the equation of state of magnetized quark matter [Phys. Rev. C 91, 065205 (2015)]}

\author{D\'{e}bora P. Menezes}
\affiliation{Depto de F\'{\i}sica - CFM - Universidade Federal de Santa
Catarina  Florian\'opolis - SC  - CEP 88.040 - 900 - Brazil}
\author{Marcus B. Pinto}
\affiliation{
Depto de F\'{\i}sica - CFM - Universidade Federal de Santa
Catarina  Florian\'{o}polis - SC -  CEP 88.040 - 900 - Brazil}
\author{Constan\c ca Provid\^encia}
\affiliation{CFisUC, Department of Physics,
University of Coimbra, P 3004-516  Coimbra, Portugal}


\maketitle 

In this erratum we present the correct result for the magnetization
of the three flavor Nambu--Jona-Lasinio model which was originally
obtained by us in Ref. \cite {nos}. Unfortunately,   the derivatives
of the condensates with respect to the magnetic field,
$\phi_f^\prime$, which do not contribute due the extremization
conditions ($\partial \Omega/ \partial \phi_f=0$), were mistakenly
considered in Eq. (3.4) therein. Once  this mistake is corrected the
relation for the magnetization reads
\begin{equation}
{\cal M} = \sum_f \left ( \theta^{\prime \, mag}_f + \theta^{\prime \, med}_f \right ) \;,
\label{eq:mag}
 \end{equation}
 where $\theta^{\prime \, mag}_f$ and  $\theta^{\prime \, med}_f$ are
 presented respectively in Eqs. (A2) and (A4) of the Appendix in
 Ref. \cite {nos} . Therefore, only the kinetic term
 of the thermodynamical potential contributes to the  magnetization. This expression is consistent with the one previously obtained in   Ref. \cite{farias}. 
Using Eq. (\ref{eq:mag}) we  next reobtain the magnetization as well
as the pressure which are respectively shown in  Figs. 1 and 2 for the
same cases presented in Figs. 2 and 3 of the original paper \cite{nos}
and draw some conclusions.

We first comment on Fig. 1. Contrary to our previous findings, once
the correct expression for the magnetization, Eq. (\ref {eq:mag}), is
considered one observes that the magnetization presents much smaller  oscillations than the ones found in Ref. \cite {nos}.  Besides,  the
average 
magnetization increases   with the
increase of the magnetic field.  The results are now very similar to the
ones obtained in \cite{huang} for strange quark matter and in
\cite{debora2} for a gas of protons when the anomalous magnetic
moment is not included. The magnetization is essentially positive for
fields  $B>5\times 10^{17}$G, but below that field it may get negative values in some
restricted density regions (never lower than -0.002 fm$^{-2}$). It
remains possible to identify the superposition of the contributions
from the different quarks, which due to
their differences in mass and charge, give rise to the irregular
oscillations of the magnetization. However, the onset of the strange quark does
not produce the strong effects calculated in \cite{nos}, which was
due to the derivative of the condensate, in particular, eq. (A7) of
\cite{nos}.

The new results presented in Fig. 2 are also very much in line with the
findings of Refs \cite{huang,cp, debora1} however, different from the ones of  Ref. \cite{farias}. For fields $B\le 10^{18}$ G
 the perpendicular pressure falls on top of the parallel contribution.  $P_\perp$ differs from $P_\parallel$ only  above
$B>10^{18}$G, and the discontinuities present in the previous version
are no longer present. As a concluding remark, we can say that
the anisotropic effects on the pressure become non negligible for magnetic
 fields larger than $4 \times 10^{18}$ G, as already pointed out in \cite{debora1}.

\begin{figure}[tbp]
\centering
\begin{tabular}{cc}
\subfloat[]{\includegraphics[width=0.5\linewidth]{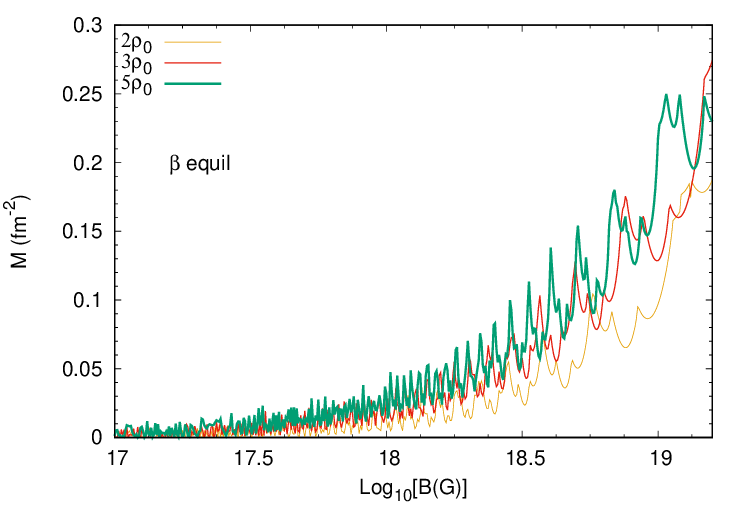} \label{fig2:a}}&
\subfloat[]{\includegraphics[width=0.5\linewidth]{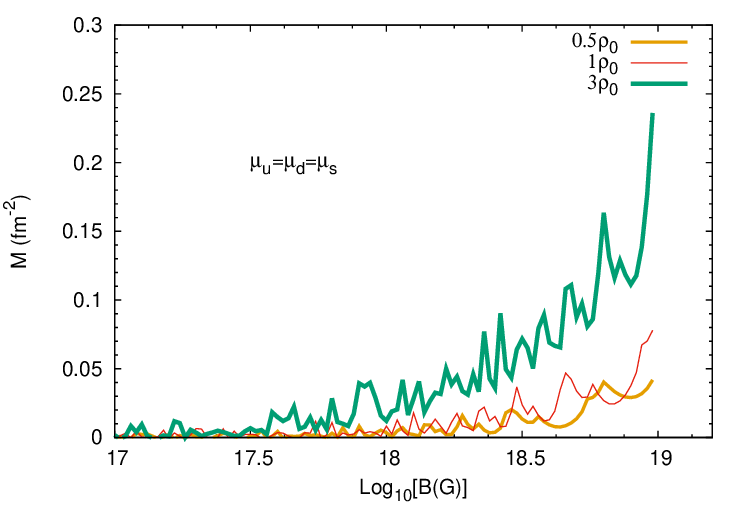} \label{fig2:b}}\\
\subfloat[]{\includegraphics[width=0.5\linewidth]{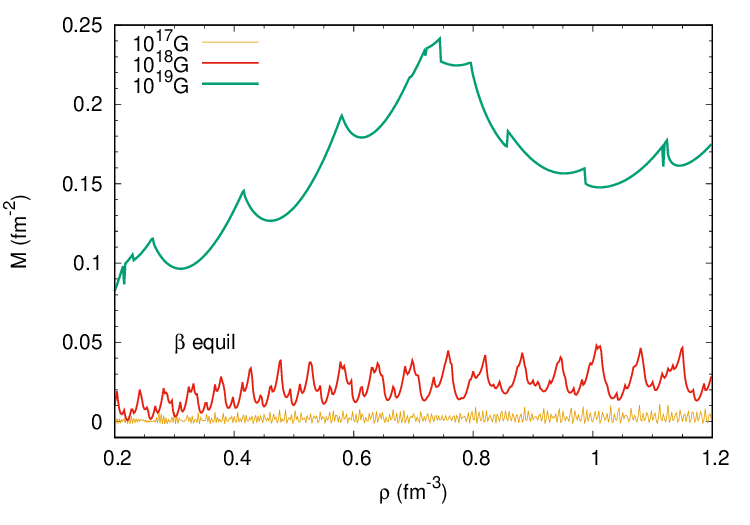} \label{fig2:c}}&
\subfloat[]{\includegraphics[width=0.5\linewidth]{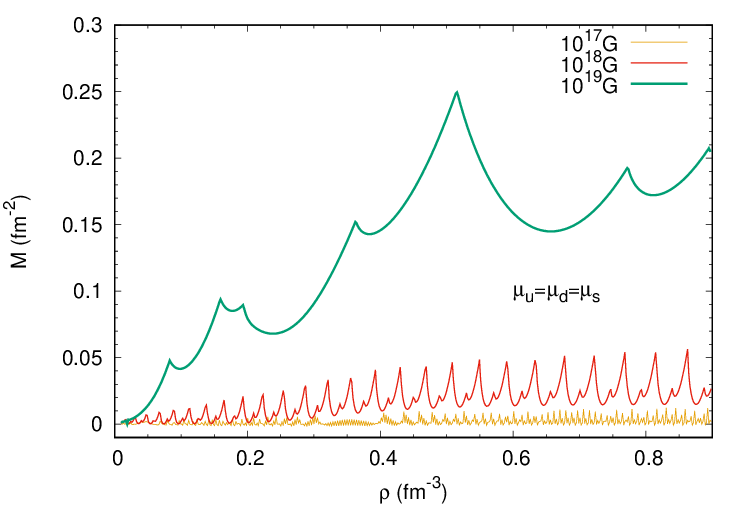} \label{fig2:d}}
\end{tabular}
\caption{ Magnetization for
 $\beta$-equilibrium matter (panels a and c) and isochemical potential quark
 matter (panels b and d). Densities of interest for each scenario are considered. In a)
 and b) the magnetization is shown as a function of $B$ and in c) and
 d) as a function of the baryonic density.
\label{fig1} }
\end{figure}

\begin{figure}[tbp]
\centering
\begin{tabular}{cc}
\subfloat[]{\includegraphics[width=0.5\linewidth]{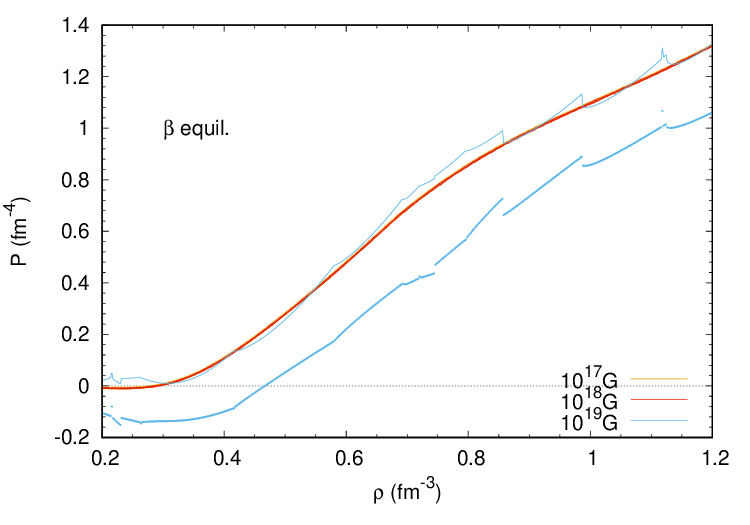} \label{fig3:a}}&
\subfloat[]{\includegraphics[width=0.5\linewidth]{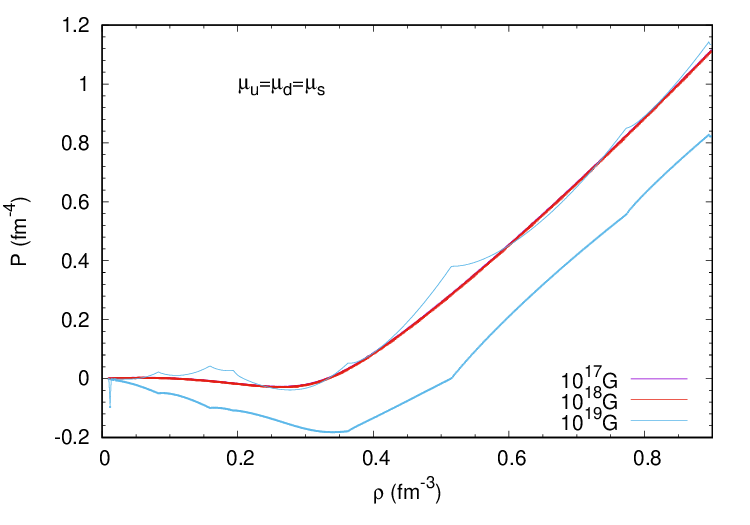} \label{fig3:b}}\\
\subfloat[]{\includegraphics[width=0.5\linewidth]{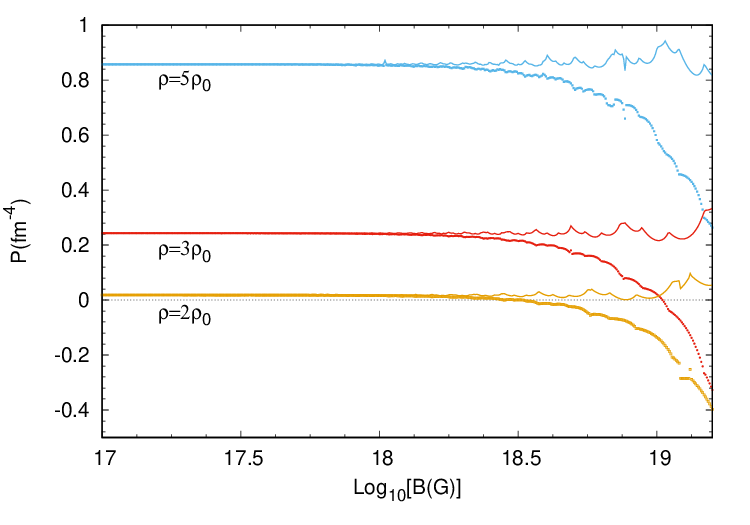} \label{fig3:c}}&
\subfloat[]{\includegraphics[width=0.5\linewidth]{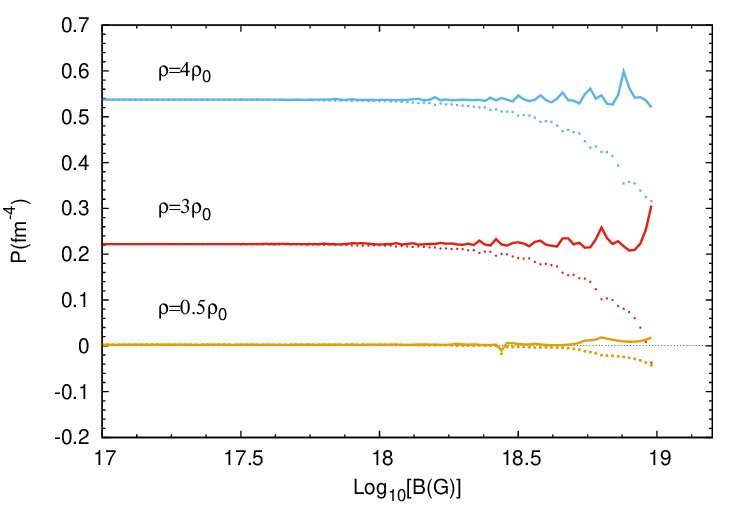} \label{fig3:d}}
\end{tabular}
\caption{Parallel (continuous lines)  and perpendicular (marks) pressures for
 $\beta$-equilibrium matter (panels a and c) and isochemical potential quark
 matter (panels b and d). Densities of interest for each scenario are considered. In a)
 and b) the pressures are shown as a function of the baryonic density
 and in c) and d) as a function of $B$.
\label{fig2}}
\end{figure}

\acknowledgements
 D.P.M. and M.B.P. are partially supported by Conselho Nacional de
Desenvolvimento Cient\'{\i}fico e Tecnol\'{o}gico (CNPq-Brazil) and  
by  Project INCT-FNA Proc. No. 464898/2014-5. C.P. acknowledges financial support 
 by Funda\c c\~ao
para a Ci\^encia e Tecnologia (FCT), Portugal, under the project
No. UID/FIS/04564/2016, and the COST action MP1304 ``NewComsptar''.